\documentclass[aps,twocolumn,showpacs,superscriptaddress]{revtex4}
\usepackage{amsfonts}
\usepackage{amsmath}
\usepackage{amssymb}
\usepackage{graphicx}

\begin{document}

\title{Matter-wave solitons and finite-amplitude Bloch waves in optical
lattices with a spatially modulated nonlinearity}
\author{Jie-Fang Zhang}
\affiliation{Institute of Nonlinear Physics, Zhejiang Normal University, Jinhua, Zhejiang
321004, P. R. China}
\author{Yi-Shen Li}
\affiliation{Department of Mathematics, University of Science and Technology of China,
Hefei, Anhui 230026, P. R. China}
\author{Jianping Meng}
\affiliation{Institute of Nonlinear Physics, Zhejiang Normal University, Jinhua, Zhejiang
321004, P. R. China}
\author{Lei Wu}
\affiliation{Institute of Nonlinear Physics, Zhejiang Normal
University, Jinhua, Zhejiang 321004, P. R. China}
\author{Boris A. Malomed}
\affiliation{Department of Physical Electronics, Faculty of Engineering, Tel Aviv
University, Tel Aviv 69978, Israel}

\begin{abstract}
We investigate solitons and nonlinear Bloch waves in Bose-Einstein
condensates trapped in optical lattices. By introducing specially
designed localized profiles of the spatial modulation of the
attractive nonlinearity, we construct an infinite number of exact
soliton solutions in terms of the Mathieu and elliptic functions,
with the chemical potential belonging to the semi-infinite bandgap
of the optical-lattice-induced spectrum. Starting from the exact
solutions, we employ the relaxation method to construct generic
families of soliton solutions in a numerical form. The stability
of the solitons is investigated through the computation of the
eigenvalues for small perturbations, and also by direct
simulations. Finally, we demonstrate a virtually exact (in the
numerical sense) composition relation between nonlinear Bloch
waves and solitons.
\end{abstract}

\maketitle

\section{Introduction}

Analogies between the electron dynamics in perfect crystals and light
propagation in periodic optical media suggest a variety of physical
phenomena and related applications. Bose-Einstein condensates (BECs) in
optical lattices (OLs) not only represent an ideal tool for investigating
fundamental effects, such as the Landau-Zener tunneling, Josephson
oscillations, dynamical instabilities, and quantum phase transitions between
the superfluidity and the Mott insulator, but also offer versatile setups
for the potential implementation of quantum computation schemes \cite%
{rmp,review}.

The mean-field description of the BEC dynamics at zero temperature is based
on the Gross-Pitaevskii equation (GPE), that is, the nonlinear Schr\"{o}%
dinger equation (NLSE) with a potential term, which is a ubiquitous model
with important realizations in other fields -- first of all, nonlinear
optics \cite{agrawal}. Many experimental and theoretical works \cite%
{anderson,12,13,14,15,16,17,18,19,20,ldcarr,gang,yang} (see also reviews
\cite{Konotop,rmp,review,Kevrekidis}) have been dealing with matter-wave and
optical solitons in OLs. Usually, these solitons are found in a numerical
form, with their chemical potential falling into bandgaps of the spectrum
induced by the OL potential, in the framework of the corresponding linear
Schr\"{o}dinger equation. A specific dynamical phenomenon, which is relevant
to the present work, is the composition relation between nonlinear Bloch
waves (NBWs) and fundamental gap solitons, whose main peaks are confined to
a single OL cell \cite{biao}.

Current experiments with BECs wield a high degree of control over key
parameters of the system. By means of the Feshbach resonance-technique,
driven by magnetic or optical fields \cite{mag_fesh,opt_fesh}, one can
adjust almost at will the strength and sign of the inter-atomic interaction.
On the other hand, available fabrication technologies allow a modulation of
nonlinearity in nonlinear optics. Therefore, there has been increased
interest in the study of the nonlinear dynamics under spatially modulated
nonlinearities, in optics and BEC alike, see original works \cite%
{Isaac,Caputo,Theocharis,Perez,jbb,jbb2,ep1,boris,boris2,Abdullaev,Greece,2D,icfo1,HK}
and book \cite{book}. In such settings, the nonlinear dynamics
exhibits novel features, such as the ``anti-Vakhitov-Kolokolov"
criterion which controls the stability of gap solitons in media
combining a spatially periodic nonlinearity and the OL potential
\cite{boris2}.

Exact solutions for matter-wave solitons in BECs with OL potentials are
important not only because of their simplicity and the connection to
physical bound states, but also since they can be used to test various
approximate methods, and may also find applications in other fields. The
objective of the present work is to construct one-dimensional soliton
solutions in physically relevant situations combining the OL potential and a
spatially modulated attractive nonlinearity. In addition to producing exact
soliton solutions in specially devised versions of such systems and
exhibiting their relation to NBWs, we also find generic numerical solutions,
by means the relaxation method, and investigate their stability. The results
may be also be directly applied to nonlinear optical media with embedded
periodic gratings, which play the same role in photonics as the OLs in BEC.

\section{The model and its reduction}

We consider a condensate of atoms trapped by a combination of a tight
cigar-shaped magnetic trap and an OL potential acting in the longitudinal
direction. If the transverse dimensions are comparable to the healing
length, and the longitudinal dimension is much longer than the transverse
ones, the setting is effectively one-dimensional, obeying by the respective
version of the GPE (see, e.g., Refs. \cite{ldcarr}):%
\begin{equation}
i\psi _{t}=-\psi _{xx}+[2V_{0}\cos (2x)+g(x)|\psi |^{2}]\psi ,
\label{original0}
\end{equation}%
where $\psi (x,t)$ is the macroscopic wave function of the condensate. Here,
time $t$, spatial coordinates $x$, and the strength of the OL potential, $%
V_{0}$, are normalized, respectively, by $\hbar /E_{r}$, $k$, and $E_{r}/4$,
with the recoil energy $E_{r}=\hbar ^{2}k^{2}/2m$, wave number of the
optical lattice $k$, and atomic mass $m$. The nonlinearity coefficient is $%
g=4m\omega _{r}a_{s}/\hbar k^{2}$, where $\omega _{r}$ is the
transverse harmonic frequency and $a_{s}$ is the $s$-wave scattering
length of inter-atomic collisions. By means of the
Feshbach-resonance technique controlled by properly designed
configurations of external fields, $a_{s}$ may be subject to a
spatial modulation, hence the corresponding nonlinearity
coefficient, $g(x)$, may be a function of $x$. In this paper, we
focus on the attractive nonlinearity, namely, $g(x)<0$, rather than
more general situations with the sign-changing $g(x)$, such as those
considered in some other works (see, in particular, Refs.
\cite{boris,boris2,HK}). It is relevant to mention that the cubic
nonlinearity in Eq. (\ref{original0}) is valid if the density is
small enough; otherwise, the reduction of the
dimension in the GPE from 3 to 1 leads to a nonpolynomial nonlinearity \cite%
{nonpoly}.

Stationary soliton solutions to Eq. (\ref{original0}) are searched
as $\psi (t,x)=\phi (x)\exp (-i\mu t)$, where chemical potential
$\mu $ is normalized by the recoil energy, and real function $\phi
(x)$ obeys the following stationary NLSE,
\begin{equation}
\mu \phi =-\phi _{xx}+[2V_{0}\cos (2x)+g(x)\phi ^{2}]\phi ,  \label{original}
\end{equation}%
with boundary conditions $\phi (x\rightarrow \pm \infty )=0$. Up to the
rescaling, the number of atoms and energy of the localized state are
\begin{equation}
N=\int_{-\infty }^{\infty }\phi ^{2}dx,  \label{atoms}
\end{equation}

\begin{equation}
\begin{split}
E_{n}& =\int_{-\infty }^{\infty }\left[ \left( \frac{\partial \phi }{%
\partial x}\right) ^{2}+2V_{0}\cos (2x)\phi ^{2}+\frac{g(x)}{2}\phi ^{4}%
\right] dx \\
& \equiv \mu N-\frac{g(x)}{2}\int_{-\infty }^{\infty }\phi
^{4}dx.
\end{split}
\label{energy}
\end{equation}

Following the scheme proposed in Refs. \cite{jbb,jbb2}, \emph{exact} soliton
solutions can be constructed by casting Eq. (\ref{original}) into the from
of a \emph{solvable} stationary NLSE in the free space,
\begin{equation}
EU=-U_{XX}+g_{0}U^{3},  \label{solvable}
\end{equation}%
where $E$ and $g_{0}$ are constants. This reduction may be implemented by
employing the transformation,
\begin{equation}
\phi (x)=\rho (x)U[X(x)],\ \ X(x)\equiv \int_{0}^{x}\rho (s)^{-2}{ds},
\label{transformation}
\end{equation}%
and requiring
\begin{equation}
g(x)=g_{0}\rho^{-6} (x),  \label{g}
\end{equation}%
where $\rho $ obeys the \textit{Ermakov-Pinney equation} \cite%
{jbb,jbb2,ep1,ep2},
\begin{equation}
\rho _{xx}+[\mu -2V_{0}\cos (2x)]\rho =E\rho ^{-3}.  \label{rho}
\end{equation}

It is commonly known that Eq. (\ref{solvable}) possesses exact solutions in
terms of the Jacobi's elliptic functions. Therefore, exact soliton solutions
to Eq. (\ref{original}) can be constructed as long as exact solutions of Eq.
(\ref{rho}) are known. In fact,
\begin{equation}
\rho =\sqrt{\alpha \varphi _{1}^{2}+2\beta \varphi _{1}\varphi _{2}+\gamma
\varphi _{2}^{2}},  \label{rho_exact}
\end{equation}%
solves the Ermakov-Pinney equation, where $\alpha $, $\beta $ and $\gamma $
are real constants satisfying $E=(\alpha \gamma -\beta ^{2})$, $\varphi _{1}=%
\mathrm{MathieuC}(\mu ,V_{0},x)$ and $\varphi _{2}=\mathrm{MathieuS}(\mu
,V_{0},x)$ are two linearly independent Mathieu functions that satisfy the
Mathieu equation \cite{mathieu1,mathieu2}, $\varphi _{xx}+[\mu -2V_{0}\cos
(2x)]\varphi =0$. For the soliton solutions to be physical, from Eq. (\ref{g}%
) it follows that $\rho (x)$ must not change its sign at any point (a
sign-definite function), otherwise the local nonlinearity would diverge at
points of $\rho =0$. Therefore, parameters $\alpha $, $\beta $, $\gamma $ in
Eq. (\ref{rho_exact}) should be chosen so as to secure this condition.

\section{Exact soliton solutions with the attractive nonlinearity}

For the attractive nonlinearity, $g_{0}<0$, a relevant exact nontrivial
solution to Eq. (\ref{solvable}) is
\begin{equation}
U(X)=\sqrt{(E-\lambda ^{2})/{g_{0}}}{\rm cn}(\lambda X-X_{0},m),
\label{cn}
\end{equation}%
where $\lambda $ and $X_{0}$ are two arbitrary constants, $E$ satisfies $%
-\lambda ^{2}\leq E<\lambda ^{2}$, and $\mathrm{cn}$ is the
Jacobi's elliptic function with module $m=\sqrt{(\lambda
^{2}-E)/{2\lambda ^{2}}}$. When $|E|<\lambda ^{2}$, Eq. (\ref{cn})
gives a periodic function of $X$, with the minimum period
$4K(m)/|\lambda|$, where $K(m)$ is the complete elliptic integral
of the first kind.
Since $\rho (x)\neq 0$, the boundary condition $\phi (x\rightarrow
\pm \infty )=0$ is satisfied when $U(X(x\rightarrow \pm \infty
))=0$. According to the periodicity of Jacobi $\mathrm{cn}$
function, $\lambda \lbrack X(x\rightarrow +\infty
)-X(x\rightarrow -\infty )]$ should be $2nK(m)$, with integer $n$. When $%
E=-\lambda ^{2}$, solution (\ref{cn}) goes over into the well-known
elementary one, $U(X)=\sqrt{{2E}/{g_{0}}}{\rm sech}\left( \sqrt{-E}%
X-X_{0}\right) $, where the boundary condition, $\phi (x\rightarrow \pm
\infty )=0$, may be satisfied if $X(x\rightarrow \pm \infty )=\infty $.

\subsection{The case of $E=0$}

For $E=0$, Eq. (\ref{rho}) is linear, and its solution can be a linear
combination of the Mathieu functions,
\begin{equation}
\rho =c_{1}{\rm MathieuC}(\mu ,V_{0},x)+c_{2}{\rm MathieuS}(\mu
,V_{0},x), \label{rho_s}
\end{equation}%
where the constants $c_{1}$ and $c_{2}$ should be chosen so as to make $\rho
\left( x\right) $ sign-definite.

We begin by constructing exact symmetric and anti-symmetric soliton
solutions for Eq. (\ref{original}), where the spatial modulation of the
nonlinearity should be represented by an even function $\rho (x)$. Without
the loss of generality, we then set $c_{1}=1$ and $c_{2}=0$ in Eq. (\ref%
{rho_s}), hence $\rho $ is an even function of $x$. Since $\rho (x)$ should
also be a sign-definite function, chemical potential $\mu $ cannot be
arbitrary for fixed strength $V_{0}$ of the OL potential. It can then be
shown that, for given $V_{0}$, there is a cutoff value of the chemical
potential, namely,
\begin{equation}
\mu _{\mathrm{co}}\equiv \mathrm{MathieuA}(0,V_{0}),  \label{cutoff_s}
\end{equation}%
below which $\rho $ is sign-definite. Here $\mu _{\mathrm{co}}$ is an even
function of $V_{0}$, representing the first characteristic value of the
\textrm{MathieuC} function, so that $\mathrm{MathieuC}$ is a $2\pi $%
-periodic function of $x$. The Taylor expansion for small $V_{0}^{2}$ is $%
\mu _{\mathrm{co}}$
$=-{(1/2)V_{0}}^{2}+(7/128){V_{0}}^{4}+O({V_{0}}^{6})$, with $\mu
_{\mathrm{co}}=0$ at $V_{0}=0$. The cutoff chemical potential
versus $V_0$ is shown in Fig. 1.

Interestingly, we find that $\mu _{\mathrm{co}}$ is exactly the minimum
energy eigenvalue in the first Bloch band of the corresponding liner Schr%
\"{o}dinger equation with periodic potential $2V_{0}\cos (2x)$.
Thus, these exact soliton solutions of Eq. (\ref{original}) exist
in the semi-infinite bandgap.

\begin{figure}[t]
\includegraphics[width=8cm,height=6cm]{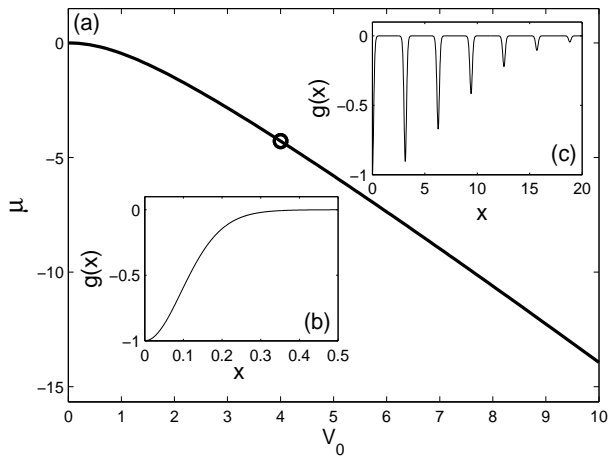}
\caption{(a) The cutoff chemical potential versus the strength of the OL
potential (solid line). Exact soliton solutions exist below the solid line.
The circle designates $\protect\mu _{\mathrm{co}}=-4.2805$ at $V_{0}=4$. (b)
The spatially-modulated nonlinearity coefficient, as given by Eq. (\protect
\ref{nonlinearity}), with $\protect\mu =-10$, $V_{0}=4$, and $g_{0}=-1$. (c)
The same as (b), except $\protect\mu =-4.2807$.}
\end{figure}

Now we investigate the properties of the nonlinearity-modulation pattern and
respective solitons. When $\mu $ is much smaller than the cutoff value $\mu
_{\mathrm{co}}$, $\rho (x)$ increases monotonically and quickly approaches
infinity. Therefore, the modulation function,
\begin{equation}
g(x)=\frac{g_{0}}{\mathrm{MathieuC}(\mu ,V_{0},x)^{6}},  \label{nonlinearity}
\end{equation}%
is localized in a very narrow single region [Fig. 1(b)]. Also, from Eqs. (%
\ref{g}) and (\ref{rho}) it follows that the smaller the chemical potential,
the narrower the localization region. On the contrary, when $\mu $
approaches $\mu _{\mathrm{co}}$, $\rho $ oscillates and slowly approaches
infinity, so that the region of the localization of $g(x)$ is relatively
wide, featuring several layers [Fig. 1(c)], and the more closely the
chemical potential approaches the cutoff value, the wider the localization
region of the nonlinearity coefficient.

Since the even and sign-definite $\rho (x)$ approaches infinity at $%
|x|\rightarrow \infty $, it is clear that $X(x)$, defined in Eq. (\ref%
{transformation}), is a monotonic non-decreasing odd function of $x$, which
has upper and lower limits. Therefore, to let the exact soliton solutions
meet the boundary condition $\phi (x\rightarrow \pm \infty )=0$, constant $%
\lambda $ in Eq. (\ref{cn}) must be chosen so as to satisfy condition $%
\lambda X(x\rightarrow +\infty )=nK(\sqrt{2}/2)$, where $n=1,2,3,\cdots $.
At the same time, constant $X_{0}$ should be chosen as $X_{0}=0$ for even
integer $n$, and $X_{0}=K(1/\sqrt{2})$ for odd integer $n$. Thus, exact
soliton solutions to Eq. (\ref{original}), with the modulation pattern taken
as per Eq. (\ref{nonlinearity}), are

\begin{equation}
\begin{split}
\phi _{n}(x)& =\frac{nK(1/\sqrt{2})}{\sqrt{-g_{0}}{X(+\infty )}} \mathrm{%
MathieuC}(\mu ,V_{0},x) \\
& \times \mathrm{cn}\left( \frac{nK(1/\sqrt{2})}{X(+\infty )}X,1/\sqrt{2}%
\right) ,
\end{split}
\label{symmetrical_sol}
\end{equation}%
for $n=1,3,5,...$ , while for $n=2,4,6,...$ the exact solutions are
\begin{equation}
\begin{split}
\phi _{n}(x)& =\frac{nK(1/\sqrt{2})}{\sqrt{-g_{0}}{X(+\infty )}} \mathrm{%
MathieuC}(\mu ,V_{0},x) \\
& \times \mathrm{cn}\left[ \frac{nK(1/\sqrt{2})}{X(+\infty )}X-K\left( \frac{%
\sqrt{2}}{2}\right) ,\frac{\sqrt{2}}{2}\right] ,
\end{split}
\label{symmetrical_sol2}
\end{equation}%
where we define $X(x)=\int_{0}^{x}\mathrm{MathieuC}(\mu ,V_{0},s)^{-2}ds.$

\begin{figure}[t]
\includegraphics[width=8cm,height=6cm]{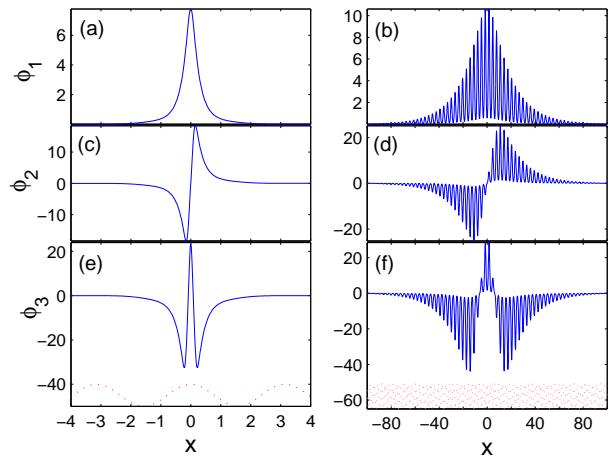}
\caption{(Color online) Exact symmetric solitons for (a) $n=1$ and (e) $n=3$%
, and an exact antisymmetric soliton for (c) $n=2$, where the corresponding
modulation function $g(x)$ is taken as per Fig. 1(b). Panels (b), (d) and
(f) are the same as (a), (c), and (e), respectively, expect the
corresponding modulation function is taken as in Fig. 1(c). Solid circles in
the bottom of each column show the OL potential.}
\end{figure}

It follows from Eqs. (\ref{transformation}), (\ref{nonlinearity}), (\ref%
{symmetrical_sol}), and (\ref{symmetrical_sol2}) that, once the chemical
potential ($\mu <\mu _{\mathrm{co}}$), constant $g_{0}$, and the strength of
the OL potential, $V_{0}$, are fixed, there exists an \emph{infinite number}
of exact solitons sharing the same chemical potential. Note that expression (%
\ref{symmetrical_sol}) is an even function of $x$, hence the soliton is
symmetric. On the contrary, expression (\ref{symmetrical_sol2}) is an odd
function of $x$, which varies $\sim x$ at $x\rightarrow 0$, yielding an
antisymmetric soliton. In either case, the matter-wave densities are even
functions of $x$. The exact soliton solution, $\phi _{n}$, possess $n-1$
density nodes [see Figs. 2 and 3], and from Eq. (\ref{energy}) it can be
concluded that the larger $n$, the larger the energy of the corresponding
BEC state. Thus one may conclude that $\phi _{1}$ corresponds to the ground
state, $\phi _{n}$ corresponding to the $(n-1)$-th excited states. By
comparing the exact soliton solution $\phi _{1}$ with the ground-state
solution of the same GPE, obtained in a numerical form by means of the
imaginary-time method, we find that $\phi _{1}$ is identical to the ground
state when $V_{0}<0$. However, $\phi _{1}$ is not always the ground state
when $V_{0}>0$ (for instance, $\phi _{1}$ remains the ground-state solution
at $\mu <-9$ for $V_{0}=4$). On the other hand, for the one-dimensional
linear Schr\"{o}dinger equation, it is well-known that localized states with
different energy eigenvalues are orthogonal. Here we find that the localized
states of the \emph{nonlinear} GPE are \emph{not} orthogonal.

\begin{figure}[t]
\includegraphics[width=7cm,height=5cm]{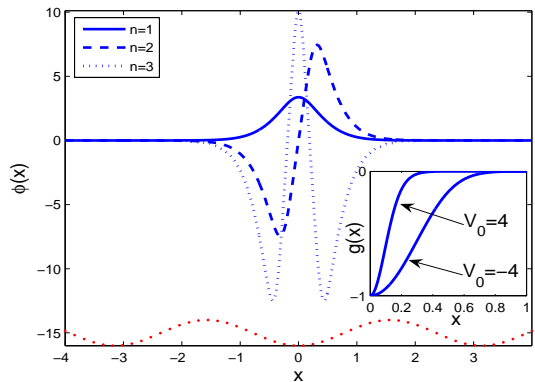}
\caption{(Color online) Exact symmetric and anti-symmetric solitons with
chemical potential $\protect\mu =-10$. The parameters are $V_{0}=-4$, $%
g_{0}=-1$. Solid circles show the OL potential.}
\end{figure}

From Figs. 1 and 2 it can be found that the widths of solitons are
proportional to the widths of the respective nonlinearity-modulation
profiles, $g(x)$. This is understandable because both the widths of the
solitons and $g(x)$ profiles are determined by $\rho (x)$, see Eqs. (\ref%
{transformation}) and (\ref{g}). That is, the more rapidly $\rho
(x)$ approaches infinity, the narrower the solitons and $g(x)$
distributions are. Further, it can be shown that the widths of the
exact solitons are always larger than those of the respective
modulation profiles. To analyze this
point in a simple form, we here take the case of $V_{0}=0$. In this case, $%
\mu <\mu _{\mathrm{co}}=0$, $\rho \sim \exp (\sqrt{-\mu }x)$, so that $g\sim
\exp (-6\sqrt{-\mu }x)$ and $R\sim 1-\exp (-2\sqrt{-\mu }x)$ at $x>0$; thus
the width of the soliton is about three times larger than that of the $\rho
(x)$ modulation. Since exact solitons in the left column of Fig. 2 are
confined mainly to a single OL cell, they can be called fundamental gap
solitons \cite{biao}, whereas the right column displays broader gap
solitons, alias gap waves \cite{gap-wave}.

Another noteworthy point is that, although the cutoff chemical potential, $%
\mu _{\mathrm{co}}$, is an even function of $V_{0}$, the
nonlinearity-modulation function is not. From Eqs. (\ref{rho}) and (\ref%
{nonlinearity}), it follows that the width of $g(x)$ corresponding to $%
V_{0}>0$ is smaller than that for $-V_{0}$, see a typical comparison in Fig.
3. Thus, the effective nonlinearity in the case of $-V_{0}$ is stronger than
for $V_{0}$, at the same $g_{0}$. On the other hand, the effective potential
is attractive (repulsive) for negative (positive) $V_{0}$ for fundamental
gap solitons. For these reasons, the number of atoms in the exact
fundamental gap solitons with $V_{0}<0$ is smaller than for $V_{0}>0$, as
shown by Figs. 2 and 3.

Similarly, exact asymmetric solitons to Eq. (\ref{original}) can be
constructed if we let $c_{1}c_{2}\neq 0$ in Eq. (\ref{rho_s}). For $\rho $
to be sign-definite, we again need $\mu <\mu _{\mathrm{co}}$, and $%
c_{1},c_{2}$ should be carefully chosen. As a generic example, we take $\mu
=-5$, $V_{0}=4$, and $c_{2}=g_{0}=-c_{1}=-1$. In such a case, $\rho (x)=%
\mathrm{MathieuC}(-5,4,x)-\mathrm{MathieuS}(-5,4,x)$, the asymmetric
modulation profile is given by Eq. (\ref{g}), and $X(x)=\int_{0}^{x}\rho
^{-2}(s)ds$. To meet the boundary conditions $\phi (x\rightarrow \pm \infty
)=0$, constants $\lambda $ and $X_{0}$ in Eq. (\ref{cn}) should satisfy
\begin{equation}
\begin{split}
\lambda \lbrack X(+\infty )-X(-\infty )]& =2nK(\sqrt{2}/{2}), \\
X_{0}& =\lambda X(-\infty )+K(\sqrt{2}/{2}),
\end{split}
\label{lambda}
\end{equation}%
where $n=1,2,3,\cdots $.

\begin{figure}[t]
\includegraphics[width=7cm,height=5cm]{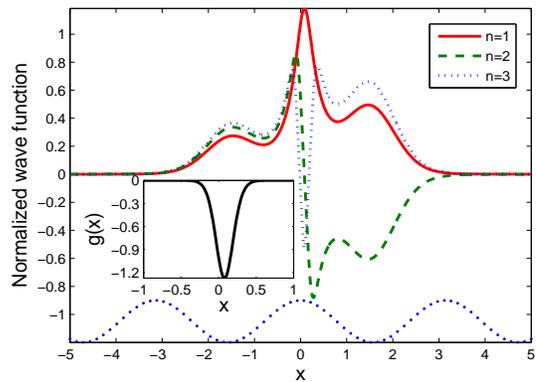}
\caption{(Color online) Normalized exact asymmetric solitons $\protect\phi %
_{n}N_{n}^{-1/2}$, where $N_{n}$ is the corresponding norm (scaled number of
atoms). Inset: the corresponding asymmetric nonlinearity-modulation profile.
Here, we set $g_{0}=-1$, with other parameters given in the text. Solid
circles show the OL potential.}
\end{figure}

The exact asymmetric solitons can be found when substituting values (\ref%
{lambda}) into Eqs. (\ref{transformation}) and (\ref{cn}). The
representative profiles of the solitons, together with the corresponding
asymmetric nonlinearity-modulation profile, are displayed in Fig. 4. Similar
to the exact symmetric and antisymmetric solutions, different solitons with
the same chemical potential are not orthogonal.

\subsection{The case of $E>0$}

The exact solution to Eq. (\ref{solvable}) with $E>0$ is given by Eq. (\ref%
{cn}). To construct the exact localized solutions, $\rho (x)$ should
approaches infinity as $|x|\rightarrow \infty $, so that the function $X(x)$
is bounded; it can be shown that such a requirement may be realized when the
chemical potential $\mu $ falls into the bandgaps of the spectrum induced by
the OL potential of the corresponding linear Schr\"{o}dinger equation. The
exact solution for $\rho $ is given by Eq. (\ref{rho_exact}). Due to the
nonzero value of the corresponding Wronskian, ${\varphi _{1}}^{\prime
}\varphi _{2}-\varphi _{1}{\varphi _{2}}^{\prime }=-1$, $\varphi _{1}(x)$
and $\varphi _{2}(x)$ are not zero at the same position, and $\rho $ is
always sign-definite. However, when $\mu $ does not belong to the
semi-infinite bandgap, there exist several points where $\rho $ is very
close to zero, making the strength of the nonlinearity very large (this
region is very narrow, and $\rho $ looks like the delta function), which we
do not consider here. We are rather interested in the case of $\mu <\mu _{%
\mathrm{co}}$.

To meet the boundary condition, we need $\lambda \lbrack X(\infty
)-X(-\infty )]=2nK(m)$, $n=1,2,3,\cdots $. Because $\lambda >\sqrt{E}$, an
inequality ensues from here,
\begin{equation}
nK\left( \frac{\lambda ^{2}-E}{2\lambda ^{2}}\right) >\frac{X(\infty
)-X(-\infty )}{2}\sqrt{E}.  \label{inequality}
\end{equation}%
From Eq. (\ref{inequality}) it follows that $n>n_{\max }\equiv
\lbrack X(\infty )-X(-\infty )]\sqrt{E}/\pi $. Thus, unlike the
case of $E=0$, where $n=1,2,3,\cdots $, here the first several
values of $n$ may disappear. For example, if $n_{\max }=2.5$, then
actual values which give rise to the solitons are $n=3,4,5,\cdots
$. However, we find that, in the semi-infinite bandgap, $n_{\max
}<1$, regardless of values of $E$, $\alpha $, $\beta $, and
$\gamma $. That is to say, there is still an \emph{infinite
number} of exact solitons sharing the same chemical potential. The
exact soliton
solutions are given by Eqs. (\ref{transformation}), (\ref{rho_exact}), and (%
\ref{cn}), with the nonlinearity given by Eq. (\ref{g}).

\subsection{The case of $E<0$}

In this case, the sign-definite $\rho $ exists when the chemical
potential is in the semi-infinite bandgap, that is, $\mu <\mu
_{\mathrm{co}}$, and the real constants $\alpha $, $\beta $, and
$\gamma $ should be carefully chosen. To meet the boundary
condition, we need $\lambda \lbrack X(\infty )-X(-\infty
)]=2nK(m)$, $n=1,2,3,\cdots $. Unlike the case of $E>0$, where
inequality (\ref{inequality}) must be satisfied, here there is no
restriction on $n$. That is to say, there is still an infinite
number of exact solitons sharing the same chemical potential.

\subsection{Discussion}

Thus far, we have demonstrated above that an infinite number of exact
soliton solutions can be constructed in the model with the OL potential,
which share the same values of the chemical potential. These solutions exist
in the semi-infinite bandgap, in accordance with the fact that families of
gap solitons [$n$, for example, in Eqs. (\ref{symmetrical_sol}) and (\ref%
{symmetrical_sol2}), denotes the family's index] can be found in the
semi-infinite bandgap when the attractive nonlinearity is spatially
homogeneous \cite{biao,kivshar}. The same model also supports gap solitons
in finite bandgaps; we are not going to discuss exact solitons in those
bandgaps because the exact spatially modulated nonlinearity mimics the delta
function, which (i) may be hard to realize in experiments, and (ii) the
corresponding profile of exact solitons are irregular.

We did not consider the repulsive nonlinearity here. The reason is that, for
the repulsive nonlinearity, $g_{0}>0$, the nontrivial solution to Eq. (\ref%
{solvable}) is $U(x)=\sqrt{2(E-\lambda ^{2})/g_{0}}{\rm sn}(\lambda X-X_{0},%
\sqrt{E/\lambda ^{2}-1})$, where $\lambda ^{2}<E<2\lambda ^{2}$. To meet the
boundary condition, we must demand $\lambda \lbrack X(\infty )-X(-\infty
)]=2nK(\sqrt{E/\lambda ^{2}-1}))$, $n=1,2,3,\cdots $, from which it follows
that $n<n_{\max }\equiv \lbrack X(\infty )-X(-\infty )]\sqrt{E}/\pi $. For
the chemical potential falling into the semi-infinite bandgap, $n_{\max }<1$%
. Therefore, there are no exact solitons in the semi-infinite bandgap for
the spatially modulated repulsive nonlinearity, just like in the case of the
spatially uniform repulsive nonlinearity \cite{rmp}.

\section{Numerically found solitons and their stability}

In Sec. III, we were able to find only discrete sets of exact soliton
solutions for the given nonlinearity. Here we consider more general
matter-wave solitons with different values of the chemical potential in the
OL potential, when the localized nonlinearity-modulation profile is fixed.
That is, we aim to find solitons in the framework of equation
\begin{equation}
\mu \phi =-\phi _{xx}+2V_{0}\cos (2x)\phi -\frac{\phi ^{3}}{\mathrm{MathieuC}%
(\mu _{0},V_{0},x)^{6}},  \label{orig}
\end{equation}%
with $\mu _{0}<$ $\mu _{\mathrm{co}}$ and, generally speaking,
$\mu \neq \mu _{0}$, where $\mu _{\mathrm{co}}$ is given by Eq.
(\ref{cutoff_s}). We focus here only on symmetric soliton
solutions in the semi-infinite bandgap, that is, $\mu<\mu_{co}$.

\begin{figure}[t]
\includegraphics[width=8cm,height=6cm]{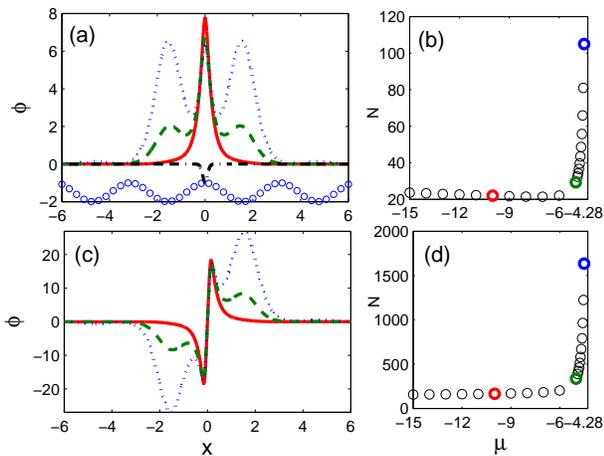}
\caption{(Color online) (a) The first family of numerically found solitons.
Solid (red), dashed (green), and dotted (blue) lines represent the solitons
with chemical potentials $\protect\mu =-10$, $\protect\mu =-5$, and $\protect%
\mu =-4.5$, respectively. The dashed-dotted line shows the
nonlinearity-modulation profile, with $\protect\mu _{0}=-10$ and $V_{0}=4$,
see Eq. (\protect\ref{orig}) (this corresponds to what is defined as \textit{%
case I} in the text). Open circles show the OL potential. (b) The number of
atoms versus the chemical potential. (c) and (d): The same as (a) and (b),
except that the solitons are from the second family.}
\end{figure}

\begin{figure}[t]
\includegraphics[width=8cm,height=6cm]{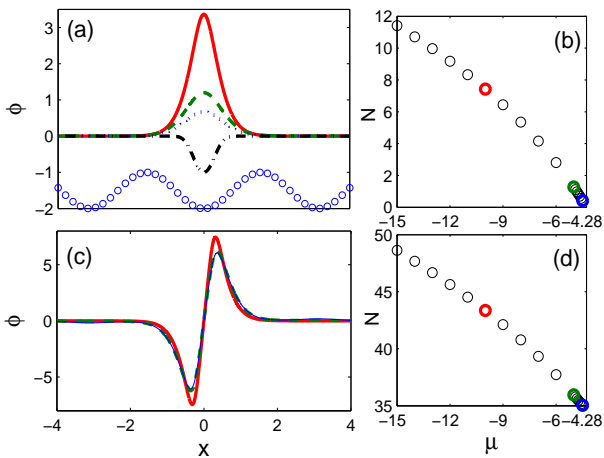}
\caption{(Color online) The same as Fig. 5, except $V_{0}=-4$ (which
corresponds to \textit{case II}, as defined in the text).}
\end{figure}

\begin{figure}[t]
\includegraphics[width=8cm,height=4cm]{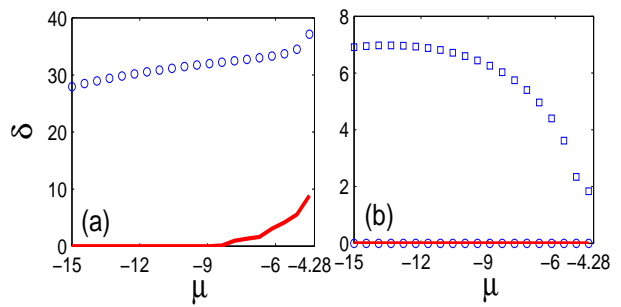}
\caption{(Color online) The largest instability growth rate versus the
chemical potential. Solid lines, circles, and squares pertain to the first,
second, and third families of the solitons, respectively. The parameters in
(a) and (b) are the same as in Figs. 5 and Fig. 6, respectively.}
\end{figure}

Exact solutions to Eq. (\ref{orig}) have been found above for $\mu =\mu _{0}$%
. Thus, using the exact solutions as an initial guess, one can find more
general solitons by means of the numerical relaxation method. There are two
different cases, which we define as I and II, with $g(x)$ localized,
respectively, around a peak or bottom of the OL potential, for positive or
negative $V_{0}$. In either case, solitons can be found in the semi-infinite
bandgap, regardless of the value of $\mu _{0}$. In case II, the number of
atoms is a monotonously decreasing function of $\mu $, just like in the case
of the NLSE with the spatially uniform attractive nonlinearity. However, the
situation is quite different in case I. For the first family solitons, we
find that the number of atoms at first decreases and then increases with the
increase of $\mu $, see Fig. 5(b), whereas for other soliton families, the
atom number is a monotonously increasing function of $\mu $, see Fig. 5(d).
These types of the dependences have obvious implications for the solitons'
stability, as per the Vakhitov-Kolokolov criterion \cite{VK}, see below. For
all soliton families in case I, when $\mu $ approaches the cutoff value,
most atoms are located in wells of the OL potential adjacent to the region
where the nonlinearity is concentrated. In other words, the solitons are
confined to one or two OL cells in case I, while in case II they are trapped
in a single cell.

Obviously, the stability of the solitons must be investigated too.
To this end, we first employ the linear-stability analysis.
Substituting a perturbed solution, $\psi (x,t)=[\phi (x)+u(x)\exp
(i\delta t)+v^{\ast }(x)\exp (-i\delta ^{\ast }t)]\exp (-i\mu t)$,
into Eq. (\ref{original0}) and linearizing it around the unperturbed
one, $\phi (x)$, we arrive at the eigenvalue problem,
\begin{equation}
\left(
\begin{array}{cc}
\mathcal{L} & -g\phi ^{2} \\
g\phi ^{2} & -\mathcal{L} \\
\end{array}%
\right) \left(
\begin{array}{c}
u \\
w \\
\end{array}%
\right) =\delta \left(
\begin{array}{c}
u \\
w \\
\end{array}%
\right) ,  \label{instability}
\end{equation}%
with operator $\mathcal{L}={d^{2}}/{dx^{2}}+\mu -2V_{0}\cos
(2x)-2g\phi ^{2}$. Here $g=-\mathrm{MathieuC}(\mu
_{0},V_{0},x)^{-6}$. The soliton is unstable if any eigenvalue
$\delta $ has an imaginary part.

\begin{figure}[t]
\includegraphics[width=8cm,height=6cm]{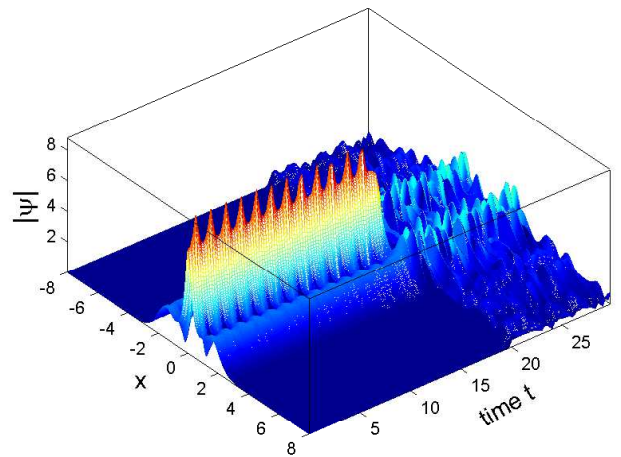}
\caption{(Color online) The evolution of an unstable soliton. The
nonlinearity-modulation function and OL potential are the same as in Fig.
5(a). The initial condition is taken as per the numerically calculated
solution [the dashed (green) line in Fig. 5(a)], mixed with a random
(white-nose) perturbation.}
\end{figure}

Results of numerical calculations displayed in Fig. 7 reveal that, in case
II, the first and second families of solitons are \emph{stable} against
small perturbations, while higher-order soliton families are unstable. On
the other hand, for case I, stable solitons emerge only in the first family,
when the chemical potential is small enough, so that the solitons are very
narrow, and the nonlinearity is strong enough to sustain solitons in the
presence of the locally repulsive OL potential. Similar conclusions
concerning the stability of solitons supported by the competing (locally
attractive/repulsive) linear and nonlinear potentials were reported in Ref.
\cite{HK}. Comparing the date displayed in Fig. 7 with panels (b) and (d) in
Figs. 5 and 6, we conclude that the Vakhitov--Kolokolov criterion ($dN/d\mu
<0$ as the necessary criterion for the stability of solitons supported by
the attractive nonlinearity \cite{VK}) holds in the present model. We have
also checked the stability of exact soliton solutions. The results are in
qualitatively agreement with those shown in Fig. 7.

The stability of the solitons was further checked by direct
numerical simulations of Eq. (\ref{original0}), producing results in
agreement with the predictions of the linear-stability analysis. In
particular, the solitons from the first family in the unstable
region originally exhibit a quasi-stable evolution and then decay,
with a larger part of the atom number located in a neighboring well
of the OL, see Fig. 8, while other unstable solitons quickly decay
into noise.

Although we have displayed here the results of the stability investigation
only for two special nonlinearities, similar conclusions hold for other
values of $\mu _{0}$ and $V_{0}$ as well. The asymmetric solitons too
demonstrate a similar behavior.

\section{The composition relation between solitons and nonlinear Bloch waves}

From Fig. 6, one can conclude find that the solitons and corresponding $g(x)$
modulation profiles are confined to a single cell. Then, it may be
interesting to form a spatially-periodic nonlinearity pattern, by placing
the same local profiles of $g(x)$ into other wells of the OL potential. In
such a case, the system may admit not only the gap solitons, but also NBWs
(nonlinear Bloch waves). For the NLSE with the spatially uniforms
nonlinearity, the intuitive concept of the NBWs built as chains of
fundamental gap solitons has been recently justified in Ref. \cite{biao},
which has produced a composition relation between NBWs and fundamental
solitons, although the relation cannot be expressed in a sufficiently simple
mathematical form.

In this section, we demonstrate that the composition relation is also
numerically valid in the GPE with the spatially periodic nonlinearity. To
this end, we consider the following periodic nonlinearity-modulation
pattern:
\begin{equation}
g_{p}(x)=\sum_{m}\frac{g_{0}}{\mathrm{MathieuC}(\mu _{0},V_{0},x-m\pi )^{6}},
\label{period_g}
\end{equation}%
where $m=0,\pm1,\pm2,\cdots$, and the summation is performed over
cells of the OL potential.

\begin{figure}[t]
\includegraphics[width=8cm,height=6cm]{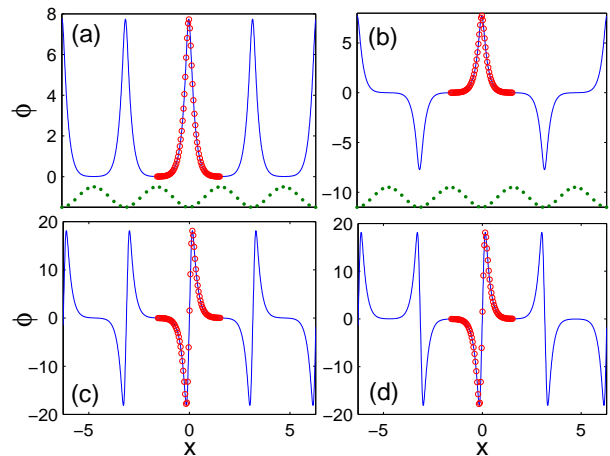}
\caption{(Color online) The composition relation between NBWs and
fundamental solitons. Opened circles in (a) and (b) represents
fundamental solitons from the first family, while in (c) and (d)
the circles denote fundamental solitons from the second family.
Solid lines in (a) and (c) are NBWs found at the center of the
Brillouin zone, while in (b) and (d) the solid lines depict the
NBWs at the edge of Brillouin zone. The profiles of the
numerically calculated NBWs completely overlap with the
expressions given by Eqs. (\protect\ref{NBW1}) and
(\protect\ref{NBW2}). The periodic nonlinearity-modulation profile
is given by Eq. (\protect\ref{period_g}), with $g_{0}=-1$,
$\protect\mu =\protect\mu _{0}=-25$, and $V_{0}=-4$. Solid circles
show the OL potential.}
\end{figure}

For many values of $\mu $, the single-peak modulation profile for $g(x)$
given by Eq. (\ref{nonlinearity}), and the respective soliton given by Eqs. (%
\ref{symmetrical_sol}) and (\ref{symmetrical_sol2}), can be confined to a
single OL cell (here we focus on the symmetric case). For instance, $g(x)$
and the soliton solution for $\mu =\mu _{0}=-25$ at $V_{0}=-4$ meet this
condition. In such cases, adjacent solitons practically do not overlap,
hence forces of the interaction between them in the periodic configuration
are negligible. Therefore, one may try to represent NBWs as chains of
fundamental solitons with identical or alternating signs:
\begin{equation}
\left( \phi _{NBW}\right) _{1}=\sum_{m}\phi _{n}(x-m\pi ),  \label{NBW1}
\end{equation}%
\begin{equation}
\left( \phi _{NBW}\right) _{2}=\sum_{m}(-1)^{m}\phi _{n}(x-m\pi ),
\label{NBW2}
\end{equation}%
where $\phi _{n}$ is given by Eq. (\ref{symmetrical_sol}) or Eq. (\ref%
{symmetrical_sol2}). The NBW described by Eq. (\ref{NBW1}) is located at the
center of the respective Brillouin zone, while that given by Eq. (\ref{NBW2}%
) is at its edge. The conjectured composition relation between the NBW and
fundamental solitons was checked numerically for the first and second
soliton families, as shown in Fig. 9. For other families of solitons, the
composition relation also holds, in the same sense.

\begin{figure}[t]
\includegraphics[width=8cm,height=6.6cm]{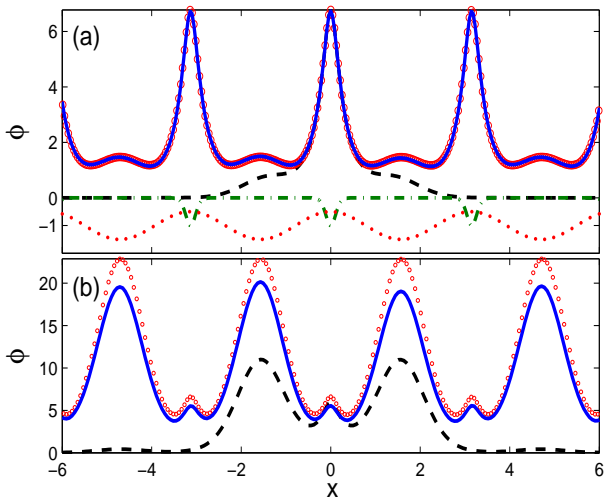}
\caption{(Color online) The composition relation between NBWs and
fundamental solitons. Dashed lines: the first family of the fundamental
solitons. Solid lines: numerically obtained NBWs at the center of Brillouin
zone. Opened circles: NBWs given by Eq. (\protect\ref{NBW1}). The periodic
nonlinearity-modulation profile, denoted by dash-dotted line in (a), is
given by Eq. (\protect\ref{period_g}) with $g_{0}=-1$, $V_{0}=4$, and $%
\protect\mu _{0}=-6$. The chemical potentials of the solitons and NBWs in
(a) and (b) are, respectively, $\protect\mu =-6$ and $\protect\mu =-4.4$.}
\end{figure}

When the soliton width exceeds the minimum periodicity of the OL, the
composition relation between NBWs and fundamental solitons, as given by Eqs.
(\ref{NBW1}) and (\ref{NBW2}), remains essentially valid, as long as the
soliton widths are smaller than two OL periods, see a typical example in
Fig. 10. In this case, in the cell where the soliton and NBW coexist, they
overlap only in the central part of the solitons. The size of the region
where the profiles of the soliton and NBW overlap is determined by the
soliton's width, hence it decreases with the increase of the width. The
composition relation effectively holds because the underlying GPE is almost
linear in the regions where the two solitons overlaps. Numerical
computations also demonstrate that one can generate confined gap waves by
putting several fundamental gap solitons together, with arbitrary
combination of signs, cf. Refs. \cite{biao,kivshar}. However, as the soliton
widths grows too large, the composition relation is no longer (numerically)
valid, see Fig. 10(b). Thus, the numerically tested composite relation
remains valid as long as the width of the individual soliton does not exceed
two OL periods (when the nonlinearity is spatially homogeneous, the
composite relation remains valid as long as the width of the individual
soliton does not exceed one OL period, as conjectured and verified in
another context in Ref. \cite{biao}).

\section{Conclusions}

We have constructed an infinite number of exact soliton solutions, both
symmetric and asymmetric, in the model of the BEC with the OL potential and
specially devised profiles of the spatial modulation of the local attractive
nonlinearity. The chemical potential of the exact solutions falls into the
semi-infinite bandgap. These solitons may coexist, with different energies,
at common values of the chemical potential.

Based on the explicit solutions, we have also found generic solitons
families in the numerical form, fixing the nonlinearity-modulation profile.
The stability of the numerically found solitons has been checked by means of
the linear-stability analysis, and also using direct simulations.

Finally, we have discussed the composition relation between nonlinear Bloch
waves and the fundamental gap solitons. We have demonstrated numerically
that the composition relation is virtually exact when widths of the solitons
do not exceed the double period of the OL. When the width of solitons
exceeds the OL period, the size of the spatial region where the solitons and
nonlinear Bloch waves overlap decreases with the increase of the soliton's
width.

\section*{Acknowledgements}

This work has been supported, in a part, by the National Natural Science
Foundation of China under Grant No. 10672147, 10704049, and 10971211, and
the Program for Innovative Research Team in Zhejiang Normal University.

\end{document}